\documentclass[prl, twocolumn, superscriptaddress,longbibliography]{revtex4-2}

\usepackage{amsmath}
\usepackage{amssymb}
\usepackage{amstext}
\usepackage{amsopn}
\usepackage{amsfonts}
\usepackage{amsxtra}
\usepackage[english]{babel}
\usepackage{graphicx}
\usepackage{bm}
\usepackage{multirow}
\usepackage{dcolumn}
\usepackage{color}
\usepackage{hyperref}
\usepackage{todonotes}
\usepackage{verbatim}
\usepackage{glossaries}

\usepackage[para]{footmisc}

\newacronym{RIXS}{RIXS}{resonant inelastic x-ray scattering}
\newacronym{CDW}{CDW}{charge-density wave}
\newacronym{SDW}{SDW}{spin-density wave}
\newacronym{QCP}{QCP}{quantum critical point}
\newacronym{ARPES}{ARPES}{angle-resolved photoemission spectroscopy}
\newacronym{APS}{APS}{Advanced Photon Source}
\newacronym{NSLS-II}{NSLS-II}{National Synchrotron Light Source II}
\newacronym{ISR}{ISR}{In-situ and Resonant}
\newacronym{MDC}{MDC}{momentum distribution curve}
\newacronym{EDC}{EDC}{energy distribution curve}
\begin{document}
 
\title{Charge density waves in cuprate superconductors beyond the critical doping}

\author{H. Miao}\email[]{hmiao@bnl.gov}
\affiliation{Condensed Matter Physics and Materials Science Department, Brookhaven National Laboratory, Upton, New York 11973, USA}
\author{G. Fabbris}
\affiliation{Advanced Photon Source, Argonne National Laboratory, Argonne, Illinois 60439, USA}
\author{R. J. Koch}
\author{D. G. Mazzone}
\altaffiliation[Present address: ]{Laboratory for Neutron Scattering and Imaging, Paul Scherrer Institut, 5232 Villigen PSI, Switzerland}
\affiliation{Condensed Matter Physics and Materials Science Department, Brookhaven National Laboratory, Upton, New York 11973, USA}

\author{C. S. Nelson}
\author{R. Acevedo-Esteves}
\affiliation{National Synchrotron Light Source II, Brookhaven National Laboratory, Upton, NY 11973, USA}

\author{G. D. Gu}
\author{Y. Li}
\affiliation{Condensed Matter Physics and Materials Science Department, Brookhaven National Laboratory, Upton, New York 11973, USA}

\author{T. Yilimaz}
\author{K. Kaznatcheev}
\author{E. Vescovo}
\affiliation{National Synchrotron Light Source II, Brookhaven National Laboratory, Upton, NY 11973, USA}

\author{M. Oda}
\author{T. Kurosawa}
\affiliation{Department of Physics, Hokkaido University, Sapporo 060-0810, Japan}

\author{N. Momono}
\affiliation{Department of Sciences and Informatics, Muroran Institute of Technology, Muroran 050-8585, Japan}

\author{T. Assefa}
\affiliation{Condensed Matter Physics and Materials Science Department, Brookhaven National Laboratory, Upton, New York 11973, USA}
\author{I. K. Robinson}
\affiliation{Condensed Matter Physics and Materials Science Department, Brookhaven National Laboratory, Upton, New York 11973, USA}
\affiliation{London Centre for Nanotechnology, University College, Gower St., London WC1E 6BT, UK}
\author{E. S. Bozin}
\author{J. M. Tranquada}
\author{P. D. Johnson}
\author{M. P. M. Dean}\email[]{mdean@bnl.gov}
\affiliation{Condensed Matter Physics and Materials Science Department, Brookhaven National Laboratory, Upton, New York 11973, USA}

\date{\today}



\date{\today}

\begin{abstract}
The unconventional normal-state properties of the cuprates are often discussed in terms of emergent electronic order that onsets below a putative critical doping of $x_\text{c}\approx0.19$. \Gls*{CDW} correlations represent one such order; however, experimental evidence for such order generally spans a limited range of doping that falls short of the critical value $x_\text{c}$, leading to questions regarding its essential relevance. Here, we use x-ray diffraction to demonstrate that \gls*{CDW} correlations in La$_{2-x}$Sr$_x$CuO$_4$ persist up to a doping of at least $x=0.21$. The correlations show strong changes through the superconducting transition, but no obvious discontinuity through $x_\text{c}\approx0.19$, despite changes in Fermi surface topology and electronic transport at this doping. These results demonstrate the interaction between \gls*{CDW}s and superconductivity even in overdoped cuprates and prompt a reconsideration of the role of \gls*{CDW} correlations in the high-temperature cuprate phase diagram. 
\end{abstract}

\maketitle

\section{Introduction}
\begin{figure*}
\includegraphics[width=\textwidth]{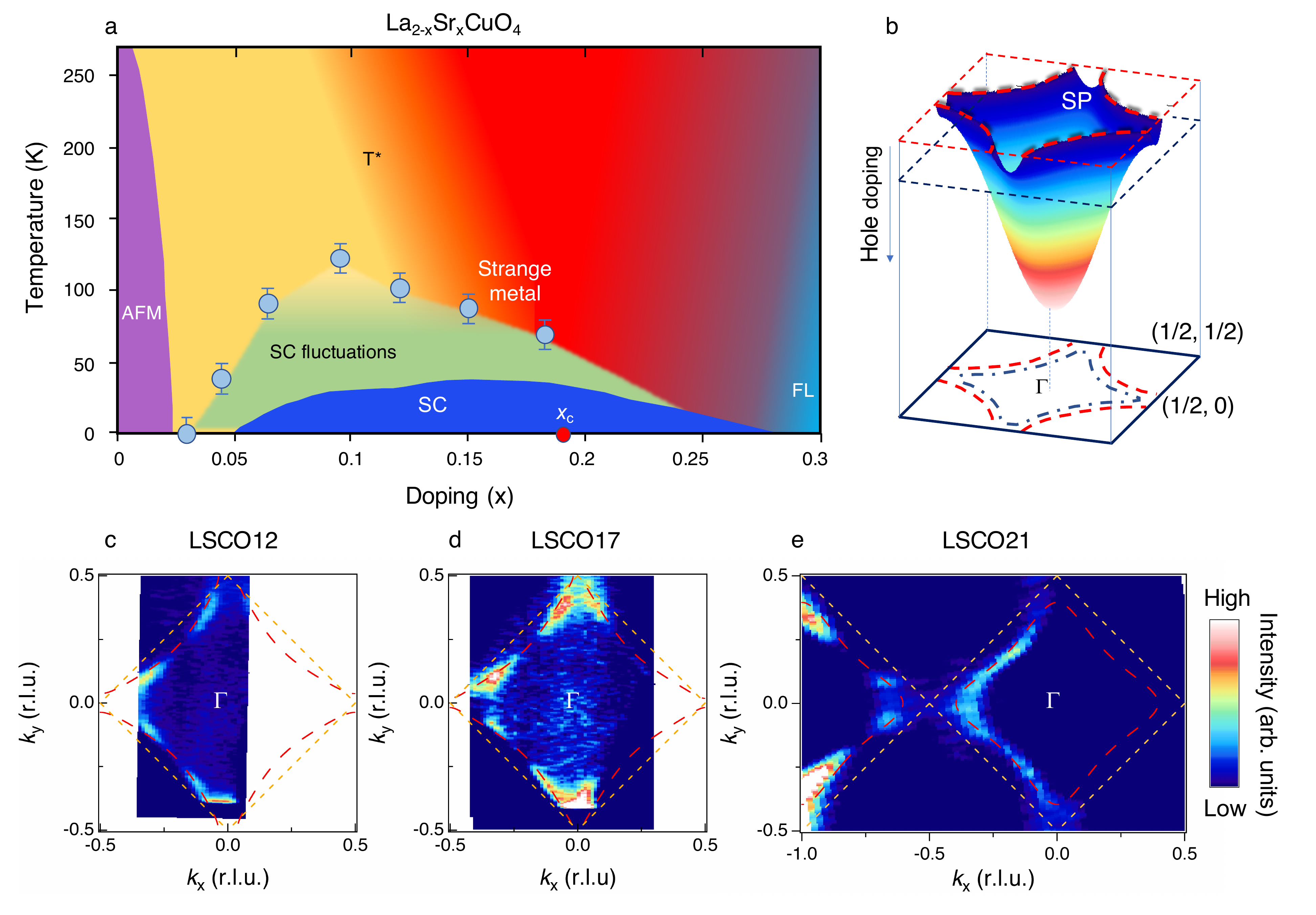}
\caption{Doping dependent electronic structure of LSCO. (a) Phase diagram of the hole-doped cuprates, constructed from magnetization, Nernst effect and resistivity data for LSCO \cite{Yamada1998, Wang2006}. $T^{*}$ is the extracted pseudogap onset temperature \cite{Wang2006,Yamada1998}. (b) Schematic band structure of LSCO. The Fermi energy, $E_F$, crosses the anti-nodal saddle-point (labeled as SP) near $x_\text{c} \approx 0.19$ triggering a Lifshitz transition. (c)-(e) Fermi surface topology of LSCO12, LSCO17 and LSCO21. The intensity plots are obtained by integrating the spectra within $\pm10$~meV of $E_F$. Orange dashes outline the antiferromagnetic Brillouin zone. Red dashed contours represent a tight-binding fit of the Fermi surface (see Supplementary note 1). The data shown in (c)-(e) were collected at 11~K.}
\label{Fig1}
\end{figure*}

The cuprate high-$T_{\text{c}}$ superconductors are often conceptualized as doped Mott insulators, in which the electronic ground state spontaneously breaks rotational and/or translational symmetry \cite{Fradkin2015, Zaanen1989, Machida1989, Emery1990}. While cuprate \gls*{CDW} correlations were discovered over two decades ago \cite{Tranquada1995}, their possible contribution to the material’s anomalous electronic properties remains a matter of vigorous debate \cite{Fradkin2015, Castellani1995, Sachdev2010, Doiron-Leyraud2012, Sebastian2015, Caprara2017}. This issue has gained increasing attention in light of the ubiquity of \gls*{CDW} order in different cuprate families \cite{Tranquada1995, Hoffman2002, Howald2003, Ghiringhelli2012, Comin2014, Tabis2014, Thampy2014, Croft2014, Christensen2014, Peng2018}. The cuprate phase diagram, shown in Fig.~\ref{Fig1}(a), shows that pseudogap, strange metal, and superconducting phases exist over an extensive doping range below a critical doping level of $x_\text{c}\approx0.19$, above which the cuprate electronic properties become gradually more Fermi-liquid-like  \cite{Wang2006, Cooper2009, Keimer2015, Ramshaw2015, Badoux2016, Giraldo-Gallo2018, Boebinger1996, Michon2019thermodynamic}. If \gls*{CDW} correlations are confined to underdoped cuprates, as previously suggested \cite{Hoffman2002, Howald2003, Ghiringhelli2012, Comin2014, Tabis2014, Thampy2014, Croft2014, Christensen2014, Peng2018, Wang2006, Cooper2009}, that would preclude the possibility of \gls*{CDW} correlations having an important role in the anomalous electronic properties. For instance, it has been argued that since \gls*{CDW} correlations disappear at $x \ll x_\text{c}$, the \gls*{QCP} at $x=x_\text{c}$ must be magnetic in nature \cite{Doiron-Leyraud2012}. Tunneling spectroscopy studies have suggested a vestigial nematic \gls*{QCP} on a similar basis \cite{Nie2014, Mukhopadhyay2019}. Very recent nuclear magnetic resonance results have reported the disappearance of spin glass behavior near $x_\text{c}$ \cite{Frachet2020hidden}. Whether this disappearance is associated with the loss of stripe correlations (i.e. coupled spin and charge density waves) remains unresolved. Moreover, the existence of \gls*{CDW} correlations is also crucial for the relevance of intertwined order. Many theoretical models for pair-density-wave superconducting states, for example, require the presence of \gls*{CDW} correlations \cite{Fradkin2015, Agterberg2020}. 

Studies of underdoped and optimally doped cuprates have shown that \gls*{CDW} correlations exist up to temperatures well above the nominal \gls*{CDW} transition temperature \cite{Miao2017, Arpaia2019}. More recently, re-entrant charge order, disconnected from the \gls*{CDW} at lower doping, was observed in overdoped {(Bi, Pb)$_{2.12}$Sr$_{1.88}$CuO$_{6+\delta}$} \cite{Peng2018}. These results motivate a reconsideration of the cuprate phase diagram, in which \gls*{CDW} correlations may extend up to higher dopings than previously thought \cite{Arpaia2019}. Herein, we address this issue by focusing on La$_{2-x}$Sr$_x$CuO$_4$ (LSCOx) ($x=0.12$, $0.17$, $0.21$, \& $0.25$) single crystals in view of its particularly well characterized transport properties and the feasibility of synthesizing high-quality samples across the entire phase diagram \cite{Cooper2009, Badoux2016, Giraldo-Gallo2018, Boebinger1996, Michon2019thermodynamic} (see Methods Section). 
%
%
\begin{figure*}
\includegraphics[width=0.9\textwidth]{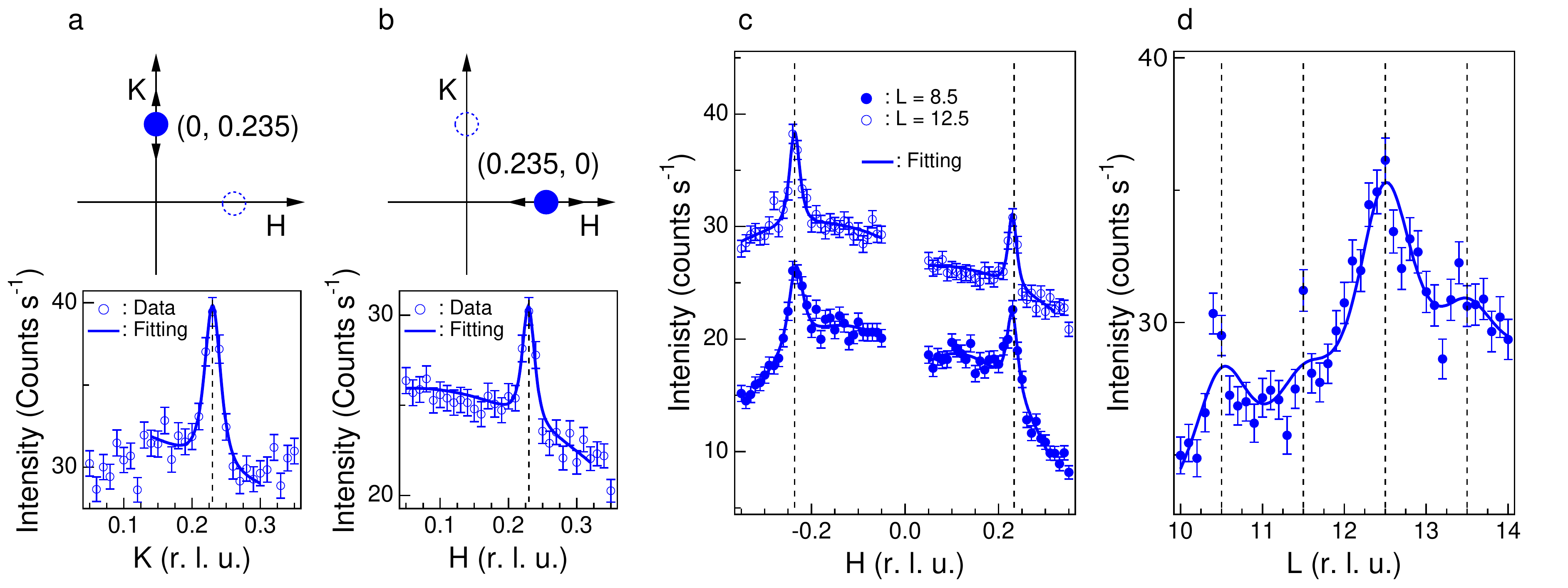}
\caption{Discovery of a \gls*{CDW} beyond $x_\text{c}$. (a) \& (b) X-ray diffraction measurements of LSCO21 at $T=16$~K along $(0,K,12.5)$ and  $(H,0,12.5)$. Supperlattice peaks are observed at $(0,0.235,12.5)$ and  $(0.235,0,12.5)$. The $H$-scans in (c) reveal further \gls*{CDW} peaks at $(\pm0.235,0,L)$ for $L=8.5$ and $12.5$. The data at $L=8.5$ are offset by $-10$~counts s$^{-1}$ for visibility. (d) The $L$-dependence of the intensity along $(-0.235,0,L)$  demonstrates poorly correlated out-of-phase \gls*{CDW} stacking  along the $c$-axis. Solid line are fits to the experimental data as described in the text and Supplementary note~2. Errorbars are one standard deviation based on Poissonian statistics.}
\label{Fig2}
\end{figure*}

\section{Results}
\subsection{Electronic structure}
Previous measurements of LSCO established the presence of a critical doping $x_\text{c}\sim 0.19$, which was defined as the doping above which the electronic transport acquires partial Fermi-liquid-like properties \cite{Cooper2009, Badoux2016}. This point coincides with, and is perhaps related to, the doping where the Fermi surface topology undergoes a Lifshitz transition \cite{Yoshida2006, Horio2018}. To prove the Sr doping, x, is consistent with previous studies, and that we indeed access the $x>x_\text{c}$ region of the overdoped phase diagram, we show the electronic structure evolution with doping in Fig.~\ref{Fig1}(b)-(e). The \gls*{ARPES} methods used are described in the Methods Section. Figure~\ref{Fig1}(b) illustrating the two-dimensional electronic structure. At low doping, to the extent that a Fermi surface exists, it is hole-like and centered at the Brillouin zone corner. With increasing hole concentration, the chemical potential drops and eventually passes through the saddle point. This results in a Lifshitz transition to an electron-like Fermi surface at the Brillouin zone center. In LSCO, the saddle point is three-dimensional with small $k_z$ dispersion. Near $x_\text{c}$, the saddle point coincidentally crosses the Fermi level. It should be noted that the carrier concentration determined by the Fermi surface (FS) area is significantly larger than the nominal Sr-doping. The origin of this effect remains, however,  unresolved \cite{Yoshida2006, Horio2018}. Nevertheless, the well-established FS evolution in LSCO can be used to confirm the Sr-doping in this study. Figure~\ref{Fig1}(c)-(e) shows \gls*{ARPES} measurements for LSCO12, LSCO17, and LSCO21. An electron-like FS is observed in LSCO21, consistent with $x_\text{c}=0.19$ and in agreement with previous \gls*{ARPES} studies \cite{Yoshida2006, Horio2018}.

\subsection{CDW order}
Having confirmed the electronic structure, we now present our main experimental finding of \gls*{CDW} correlations beyond $x_\text{c}$. Figure~\ref{Fig2} plots x-ray reciprocal space scans for LSCO21 at $T=16$~K, where reciprocal space is defined in terms of scattering vector $\bm{Q}=(H,K,L)$ using effective tetragonal lattice constants $a=b\approx3.8$~\AA{} and $c\approx13.2$~\AA{}. High sensitivity is achieved by exploiting the high brightness of the National Synchrotron Light Source II and by careful configuration of the detection system to suppress background signal (see Methods Section). Superlattice peaks are observed at $(0.235,0,12.5)$, and equivalent locations, along both the $H$ and $K$ directions [Fig.~\ref{Fig2} (a)\&(b)]. The observed $H=0.235$ matches the \gls*{CDW} wavevector in underdoped LSCO \cite{Thampy2014, Croft2014, Christensen2014, Wen2019} and is consistent with the charge stripe picture \cite{Tranquada1995}. The peaks are symmetric with respect to $\pm H$ and $K$ and are observed in multiple Brillouin zones including $(\pm 0.235,0,L)$ for $L=8.5$ \& $12.5$. An $L$-scan along $\bm{Q}=(-0.235,0,L)$ [Fig.~\ref{Fig2}(d)] reveals that the \gls*{CDW} intensity is broadly peaked at half-integer $L$ similar to underdoped LSCO \cite{Thampy2013, Croft2014, Christensen2014}. These results demonstrate the presence of \gls*{CDW} correlations beyond $x_\text{c}$. Subsequent inelastic x-ray scattering studies show that the \gls*{CDW} is associated with phonon softening even in the overdoped regime \cite{Lin2020}. 

\subsection{CDW temperature dependence \label{sec_CDW_T_dep}}
\begin{figure*}[ht]
\includegraphics[width=\textwidth]{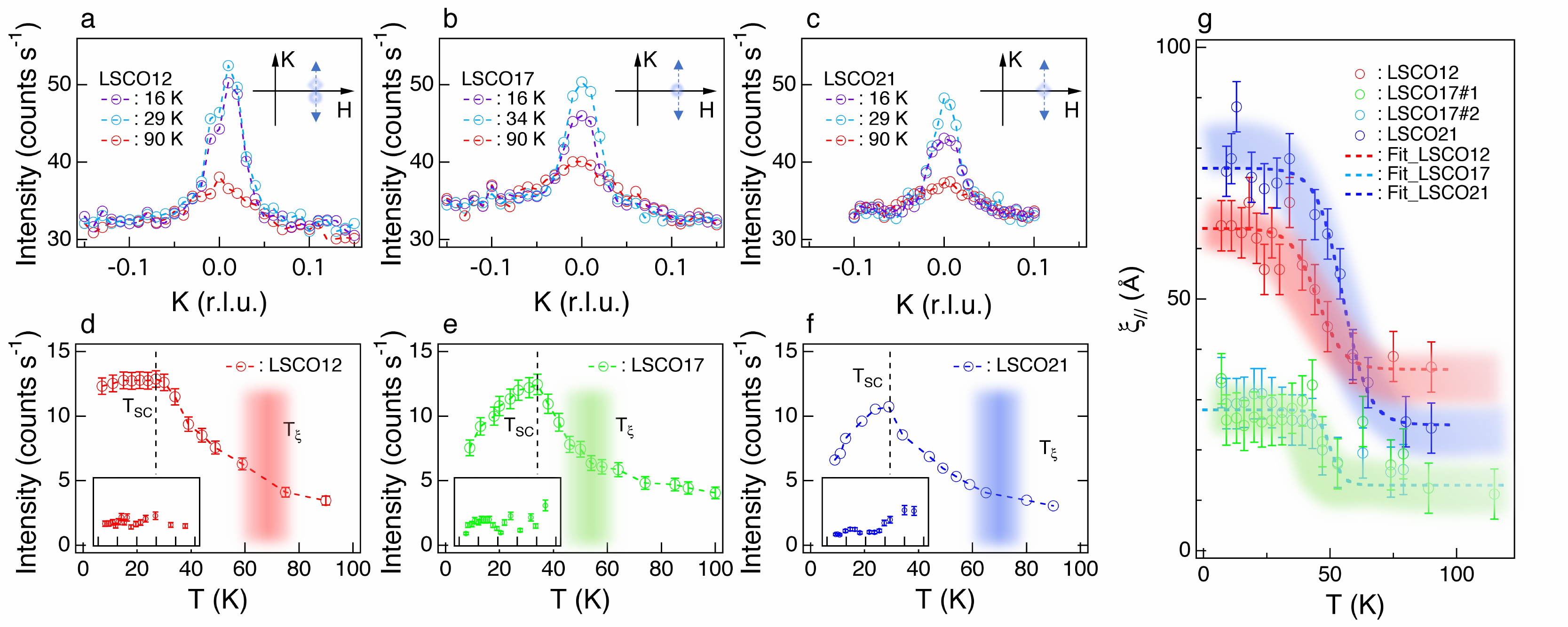}
\caption{\gls*{CDW} temperature dependence. (a)-(c) Doping dependence of the \gls*{CDW} peak intensity for temperatures $T<T_\text{SC}$, $T\approx T_\text{SC}$, and $T>T_\text{SC}$ for (a) LSCO12, (b) LSCO17, and (c) LSCO21. The inset of each panel represents the respective cut in reciprocal space. All data were taken at $L=8.5$ (d)-(f) Temperature dependence of the \gls*{CDW} intensity in LSCO for (d) LSCO12, (e) LSCO17, and (f) LSCO21. The shaded area corresponds to $T_\xi$ where the in-plane \gls*{CDW} correlation length, $\xi_\parallel$, starts to increase [as determined in (g)]. The main panels show peak height intensity and the insets show integrated intensity. (g) Temperature dependence of $\xi_\parallel$. The colored shaded curves are phenomenological fittings, $a+\frac{b}{1+e^{(T-T_{0})/4\Delta T}}$, of the temperature dependent $\xi_\parallel$ for different dopings. Here we define $T_{\xi}=T_{0}+\Delta T$. $\xi_\parallel$ increases with decreasing temperature for $T_\text{SC} <T<T_\xi$. Two independent measurements of LSCO17 samples at different beamlines show consistent suppression of $\xi_\parallel$ and $T_\xi$, indicating that systematic errors are minimal. Due to the short correlation length, the uncertainty of $T_{\xi}$ may be larger in LSCO17 than other dopings. Errorbars are one standard deviation from either Poissonian statistics or least-squares fitting.}
\label{Fig3}
\end{figure*}

Figure~\ref{Fig3} summarizes the doping and temperature dependence of the \gls*{CDW} correlations. In Fig.~\ref{Fig3}(a), Lorentzian-squared fits to the data are shown, which are parameterized in terms of amplitude, $I_{\text{CDW}}(T)$, and in-plane correlation length, $\xi_\parallel(T)=1/\text{HWHM}$ (where $\text{HWHM}$ is half-width at half-maximum) (see Supplementary note~2). Since domain formation can lead to transverse peak splitting in LSCO [c.f.~Refs.~\cite{Thampy2014, Christensen2014, Wen2019} and Fig.~\ref{Fig3}(a)], we scanned through the peaks in all three reciprocal space directions. Two Lorentzian-squared functions displaced in the $K$ (transverse) direction were used, where necessary, to account for the full intensity distribution. Peak widths and correlation lengths are determined using the $H$ (longitudinal) cut. $I_\text{CDW}(T)$ is found to be largest near $T_\text{SC}$ for all dopings [Fig.~\ref{Fig3}(b)]. Above $T_\text{SC}$, both $I_\text{CDW}$ and $\xi_\parallel(T)$ decrease with increasing temperature but remain finite up to at least $T=90$~K [Fig.~\ref{Fig3}(a)]. In agreement with previous x-ray diffraction studies of LSCO \cite{Wen2019, Thampy2014, Croft2014}, the correlation length can be separated into a marginally-ordered regime where $\xi_\parallel(T)$ is approximately 4-unit-cells (about one period of the \gls*{CDW} order), and a strongly $T$-dependent regime where $\xi_\parallel(T)$ continues to expand until superconductivity intervenes. We refer to the \gls*{CDW} in the $T$-independent regime as ``precursor'' \gls*{CDW} correlations in the sense that they come before the emergence of a stronger, more-correlated \gls*{CDW} at low temperatures. Note that for these measurements we do not have the energy resolution to directly distinguish between dynamic and static correlations. The short correlation length and quasi-temperature-independent nature of the percursor CDW indicates that it might be dynamic in nature. This phenomenology is consistent with \gls*{RIXS} scattering studies of La$_{2-x}$Ba$_x$CuO$4$ (LBCOx) and YBa$_2$Cu$_3$O$_{6+\delta}$ which show a similar two-stage \gls*{CDW} formation \cite{Miao2017, Arpaia2019, Miao2018, Miao2019}. While the \gls*{CDW} evolves smoothly from LSCO12 to LSCO21, both $\xi_\parallel$ and the onset temperature of the longer-range \gls*{CDW}, $T_\xi$, are suppressed in the overdoped regime around $x_\text{c}$ [Fig.~\ref{Fig3}(c) and Fig.~\ref{Fig4}]. The $\bm{Q}$-integrated scattering intensity, as estimated by $I_\text{CDW}\xi_\parallel^2$, shows minimal variation through $T_\text{SC}$, indicating that while superconductivity alters the  \gls*{CDW} correlation length, it does not strongly suppress the order parameter. We do not observe any \gls*{CDW} correlations in our high-sensitivity x-ray measurements at $x=0.25$ (see Supplementary note 3).

Previous measurements of the same $x=0.12$ sample allow us to compare the \gls*{CDW} order parameter, taken to be captured by the total $\bm{Q}$-integrated scattering intensity, to other cuprate systems \cite{Thampy2014}. The \gls*{CDW} order parameter of LSCO12 is only four times weaker than La$_{1.875}$Ba$_{0.125}$CuO$_4$ (which has the strongest zero-field \gls*{CDW} order). With increasing doping, the LSCO \gls*{CDW} becomes somewhat stronger for $x=0.17$ and drops appreciably for $x=0.21$ (see Supplementary note~4).  Consequently, \gls*{CDW} correlations can have an appreciable effect on the physics of LSCOx for dopings through $x_\text{c}$. 

\section{Discussion}
\begin{figure}
\includegraphics[width=\columnwidth]{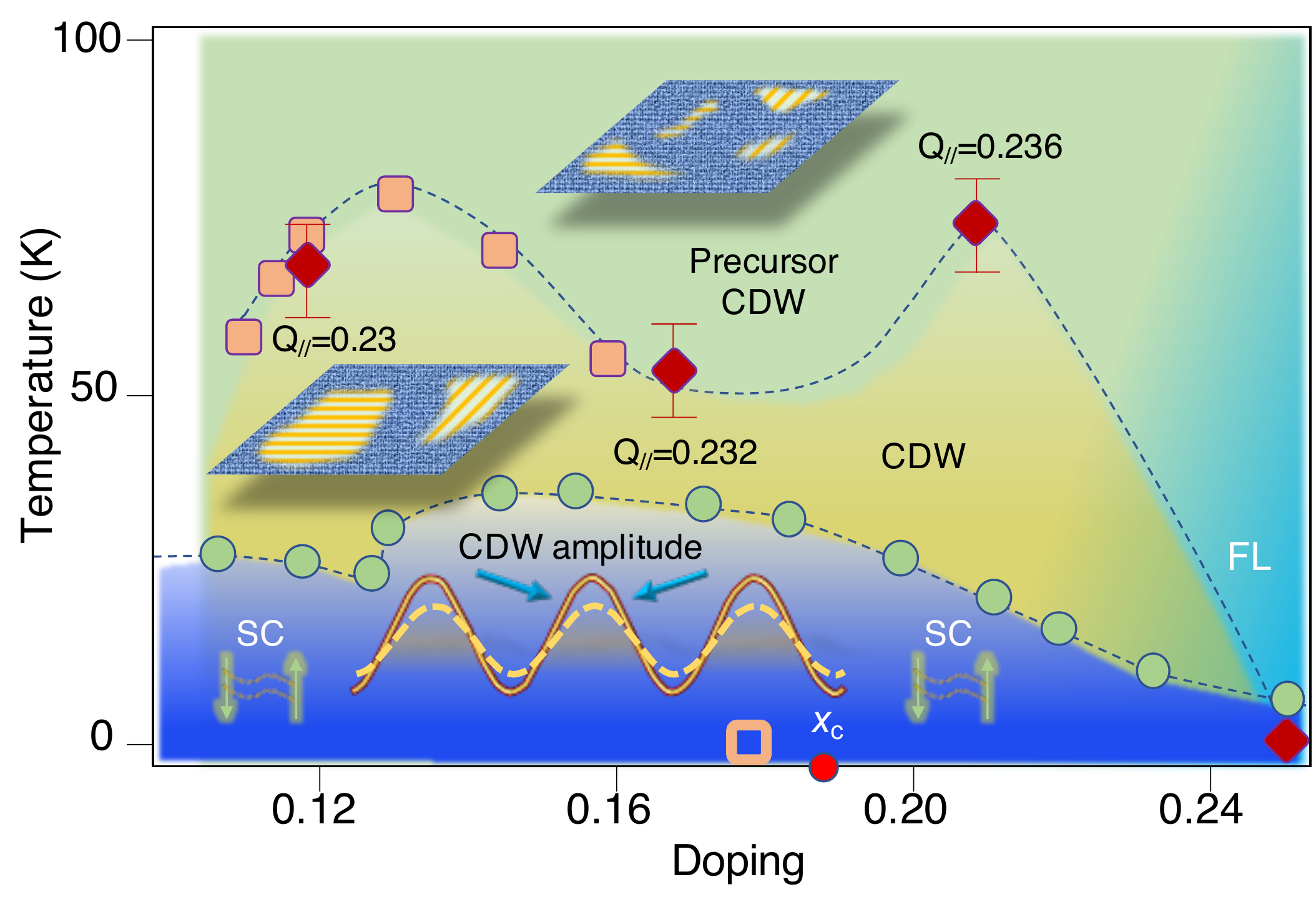}
\caption{Illustration of the extent of CDW correlations in
the cuprate phase diagram suggested by this work. Green-yellow tones represent our main result -- the presence of \gls*{CDW} correlations,   from $0.12<x<0.21$. Green denotes the precursor \gls*{CDW}, which appear at high temperature and which have a correlation length of approximately one \gls*{CDW} period \cite{Miao2017, Miao2018, Miao2019, Arpaia2019}. At lower temperature, the correlations start to grow into larger \gls*{CDW} domains, as evidenced by the increased correlation length, which we denote by the yellow tone. Red points mark where the correlation length starts to increase. This should be considered an approximate cross-over and not over-interpreted as a well-defined phase transition.  At lower temperatures still, bulk \emph{d}-wave superconductivity intervenes at $T_\text{SC}$ whereupon both the \gls*{CDW} amplitude and the correlation length saturate or start to decrease. The doping dependence reveals an anticorrelation between $T_\xi$ and $T_\text{SC}$, providing evidence for an interaction between the \gls*{CDW} and superconductivity. This is illustrated by the cartoon in the bottom of the diagram in which superconducting pairing (green spin pairs) suppress the CDW (yellow solid and dashed sinusoidal curves). The \gls*{CDW} intensity disappears in heavily overdoped LSCO25, where a Fermi-liquid-like state is recovered (Supplementary note~3 \& 5). The red diamonds reflect the present study. Pink squares and green circles are data from previous work \cite{Thampy2014, Croft2014, Wen2019, Yamada1998}.}
\label{Fig4}
\end{figure}

Figure~\ref{Fig4} summarizes our main observations -- that \gls*{CDW} correlations exist far into the overdoped regime of the cuprate phase diagram. This immediately yields three important consequences for LSCO. Firstly, very similar \gls*{CDW} properties that are observed either side of the Lifshitz transition. This provides a vivid demonstration that \gls*{CDW} correlations cannot be explained within a weak coupling Fermi surface nesting picture nor Friedel oscillations. Instead, the nearly constant $\bm{Q}_\text{CDW}$ for dopings $x\geq 0.125$ support strong coupling mechanisms, which date back to seminal work in the late 1980s \cite{Zaanen1989, Machida1989, Emery1990}. In these mechanisms one considers the balance between Coulomb interactions and kinetic energy. When doping a Mott insulators holes can save energy by clustering together as this breaks fewer magnetic bonds than widely dispersed holes. At the same time, this clustering is disfavored by the increased Coulomb repulsion and kinetic energy reduction. Since these different interactions act on different lengthscales, the overall minimum energy solution is expected to involve a spatially modulated state. Modern numerical solutions of the Hubbard model further support this idea \cite{Corboz2014, Huang2017, Zheng2017} and models based on filled stripes can reproduce a doping-independent \gls*{CDW} wavevector from $x=1/8$ to $x=1/4$ \cite{Lorenzana2002}. We also note that precursor CDW correlations are emerging as a ubiquitous feature for many cuprates, including LSCO in this study, underdoped LBCO \cite{Miao2019}, underdoped and optimally doped YBCO \cite{Arpaia2019}, underdoped Bi2212 \cite{Chaix2017} and ${\mathrm{HgBa}}_{2}{\mathrm{CuO}}_{4+\ensuremath{\delta}}$ \cite{Yu2020}. In underdoped and optimally doped YBCO, the precursor correlations appear to exist at the same wavevector around 0.3 r.l.u.\ different to the doping-dependent low-temperature CDW \cite{Arpaia2019}. It would consequently be interesting to consider a possible role for strong coupling mechanisms for all cuprates. An obviously desirable experiment would be to test whether other cuprates, such as YBCO, also exhibit CDW correlations up to similarly high dopings as LSCO. Such experiments are, however, currently held back by challenges in stabilizing high-quality heavily overdoped YBCO crystals. The robust presence of CDW correlations in LSCO seen here as a function of temperature and doping, as well as the fact that model Hamiltonian calculations reliably predict CDW correlations \cite{Corboz2014, Huang2017, Zheng2017}, would point towards their likely presence. The issue of differing wavevectors in different cuprates would, however, not necessarily be solved by such an experiment. In this regard, it is important to point out the low-temperature ordering wavevector can be influenced by coupling between the \gls*{CDW} and spin correlations or coupling between the \gls*{CDW} and the lattice, as has been suggested theoretically \cite{Zachar1998}, so differences in \gls*{CDW} wavevectors could arise from secondary interactions rather than necessarily indicating a distinct origin for the correlations. Prior work has pointed towards this as a possible explanation for temperature-induced changes in CDW wavevector in LBCO  \cite{Miao2017}.

A second immediate conclusion is that the continuous evolution of the \gls*{CDW} correlations is inconsistent with the proposed \gls*{QCP} that is associated with $x_\text{c}$ arising from \gls*{CDW} or coupled \gls*{CDW}/\gls*{SDW} order \cite{Castellani1995, Caprara2017}. Such theories can still be excluded even if one postulates a very narrow range of criticality around $x_\text{c}$, since they require either a disappearance or a symmetry change of the \gls*{CDW} through $x_\text{c}$.

Last but not least, the disappearance of CDW in LSCO25 suggests that the CDW dome in LSCO terminates between $x=0.21$ and $0.25$, where the Fermi liquid behavior starts to recover \cite{Cooper2009, Badoux2016, Giraldo-Gallo2018}. This is, again, consistent with a strong coupling CDW mechanism as Coulomb repulsion is largely screened in the Fermi liquid state. We note that in LSCO, the structural high-temperature tetragonal to low-temperature orthorhombic (LTO) phase transition  also terminates near $x=0.21$ \cite{Hucker2004}. It has been argued that the local LTO distortion may help to stabilize the CDW \cite{Chen2019}. The persistence of CDW correlations up to $x=0.21$ observed in this study is consistent with this scenario and indicates that electron-phonon coupling might be an important ingredient for the CDW formation \cite{Miao2018, Lin2020}.

The observations herin also urge a re-examination of the potential role of \gls*{CDW}s in the anomalous electronic properties of the cuprates. \gls*{CDW} correlations are a prerequisite (but not a proof) of several prominent theories of cuprate properties, which would be expected to apply across the phase diagram and not just in the underdoped region where \gls*{CDW} correlations have been studied extensively in the past. This include the possibility that \gls*{CDW} correlations play a key role in the electronic transport properties \cite{Castellani1995, Caprara2017}. Theories of pair-density-wave order \cite{Fradkin2015, Agterberg2020, Lee2014, Berg2007}, which predict competition between the \gls*{CDW} and uniform \emph{d}-wave superconductivity, also fall into this category. As shown in Fig.~\ref{Fig3}, neither the CDW peak intensity nor the CDW correlation length show the type of divergent-behavior associated with a typical phase transition. This behavior is consistent with a possible fluctuating \gls*{CDW} component, potentially influencing cuprate transport properties \cite{Badoux2016, Giraldo-Gallo2018, Boebinger1996, Michon2019thermodynamic, Arpaia2019}.

Finally, we note that a charge Bragg peak has recently been observed in overdoped (Bi,Pb)$_{2.12}$Sr$_{1.88}$CuO$_{6+\delta}$ (Bi2201), with a maximum doping comparable to that observed here \cite{Peng2018}. This state, termed re-entrant charge order, has several properties that are different to \gls*{CDW} states in LSCO and other cuprates. Re-entrant charge order appears to exist only in an isolated region of the overdoped phase diagram, disconnected from the underdoped \gls*{CDW} order. The correlation length and temperature scale of this state are also far higher than other cuprates. Intriguingly, no interaction between re-entrant charge order and superconductivity is observed in Bi2201. In contrast, similarly well-correlated \gls*{CDW} states are associated with a strong suppression of superconductivity. All these behaviors are in strong contrast with the \gls*{CDW} in overdoped LSCO, where the \gls*{CDW} wavevectors, correlation length and temperature dependence evolve smoothly from the properties of underdoped LSCO and strongly intertwine with superconductivity and low-temperature transport. Based on the electronic structure of Bi2201 and the wavevector of re-entrant charge order around 0.1~r.l.u., which extrapolates roughly linearly from the underdoped \gls*{CDW} wavevector, re-entrant charge order was proposed to arise from a van Hove singularity \cite{Peng2018}. The overdoped \gls*{CDW} in LSCO appears to have no connection to this mechanism, since the \gls*{CDW} remains unchanged regardless of the proximity to the van Hove singularity at $x = x_\text{c}$. Instead, our observations support strong coupling mechanisms.
 
In summary, high-sensitivity x-ray measurements have revealed that cuprate \gls*{CDW} correlations persist across almost the whole cuprate doping phase diagram, despite dramatic changes in the transport properties and Fermi surface topology, before disappearing when Fermi-liquid-like properties are restored. We have shown that these correlations impact superconductivity even in overdoped cuprates, suggesting that \gls*{CDW} correlations can have a far more extensive role in the cuprate phase diagram than previously envisaged, prompting investigations of \gls*{CDW} correlations in other overdoped cuprates. The discovery of \gls*{CDW}s beyond $x_\text{c}$ is confirmed by subsequent resonant inelastic x-ray scattering studies, which uncovered an unusual coupling between the CDW and lattice vibrations \cite{Lin2020}.   

\section{Methods}
\subsection{Samples}\label{sec_methods_samples}
Single crystals of La$_{2-x}$Sr$_x$CuO$_4$ ($x=0.12$, $0.17$, $0.21$ and $0.25$) were grown by the traveling-solvent floating-zone method. For each composition, a single feed rod $20-25$~cm long was used, the first few centimeters of which was removed and discarded after growth. The remaining rod was annealed in flowing O$_2$ at 980${^\circ}$C for 1~week. The superconducting transition temperatures, $28$, $37$, $30$ and $10$~K were determined by dc magnetization measurements in an applied field of 1~mT (after cooling in zero field). Our tight-binding fits to our \gls*{ARPES} measurements of these samples, described in Supplementary note~1, confirm the hole concentration matches the strontium content $x$.

\subsection{ARPES}\label{sec_methods_ARPES}
\gls*{ARPES} measurements were performed at the 21-ID-1 beamline of the \gls*{NSLS-II} using a Scienta-DA30 analyzer. Due to the small incident beam spot-size (less than $10 \times 10$~$\mathrm{\mu m}^{2}$), both the sample position and the incident light angle are fixed during the measurement. The \gls*{ARPES} intensity maps are obtained using the mapping-mode of the DA30-analyzer, which can cover 30$^{\circ}$ of cone acceptance without sample rotation. All samples were cleaved in-situ and measured at 11~K within a vacuum better than $7 \times 10^{-11}$~mbar. The photon energy was set to 60~eV for LSCO12 and LSCO17 with 18~meV energy resolution. To confirm the Fermi surface of LSCO21 is a closed loop at the $\Gamma$ point, we set the photon energy to 195~eV for LSCO21 with 90~meV energy resolution. At this energy, we were able to cover the second Brillouin zone without sample rotation. The chemical potential is calibrated based on the ARPES spectra on Silver that are recorded before and after the ARPES measurement.  

\subsection{Non-resonant hard x-ray scattering}\label{sec_methods_xray}
\begin{figure}
\includegraphics[width=\columnwidth]{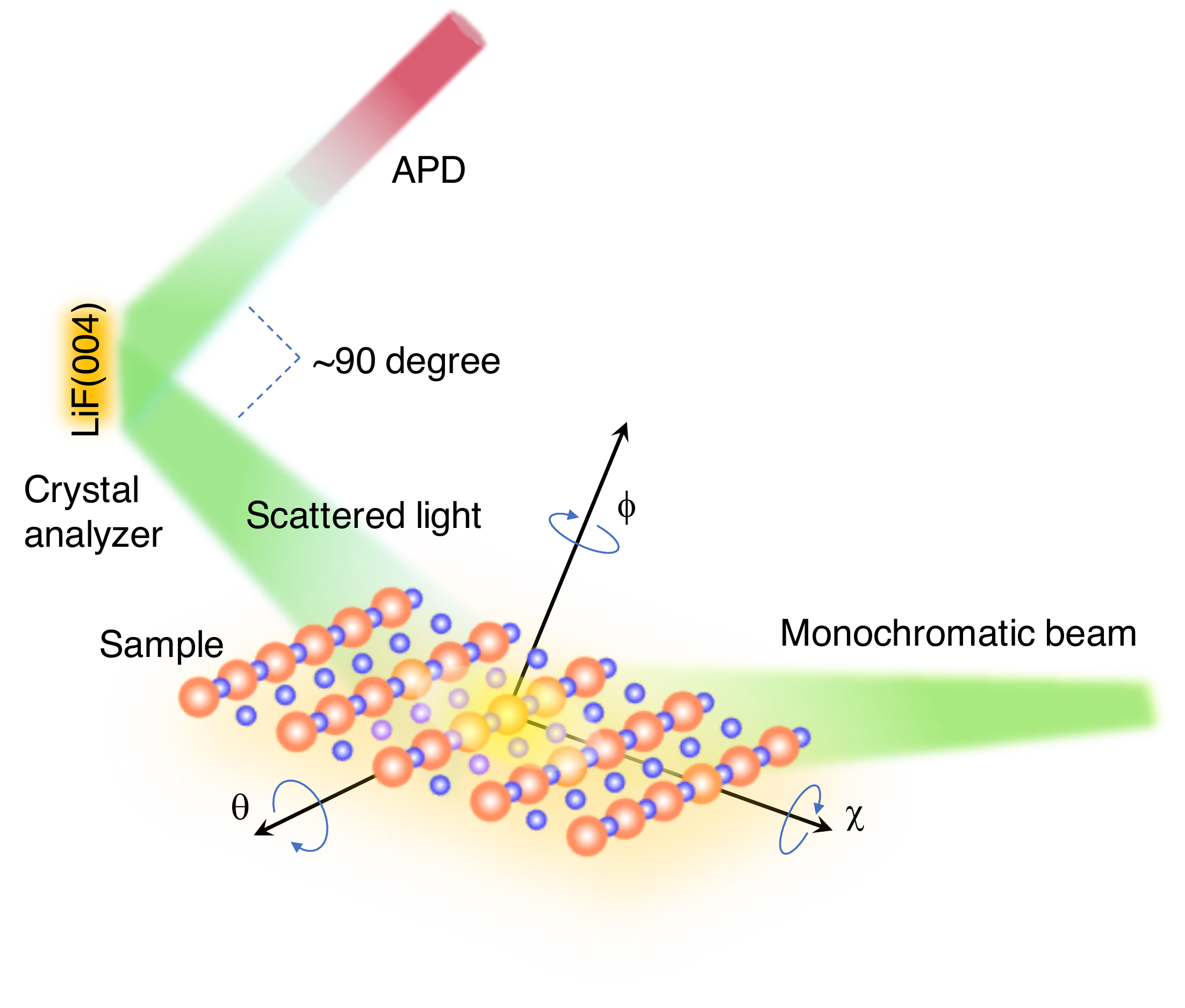}
\caption{Illustration of the experimental geometry at the 4-ID beamline of \gls*{NSLS-II}.}
\label{Fig_setup}
\end{figure}

High-precision x-ray scattering measurements were performed at the \gls*{ISR} 4-ID beamline of \gls*{NSLS-II} and 4-ID-D beamline of the \gls*{APS}. The incident photon energy was set to 8.98~keV; slightly below the Cu $K$-edge to minimize the fluorescence background. The measurements at \gls*{NSLS-II} were carried out with an avalanche photodiode (APD) detector. A LiF$(004)$ crystal analyzer was used to further suppress the background signal (Fig.~\ref{Fig_setup}). The measurements at the \gls*{APS} used a Vortex Si drift detector without any crystal analyzer.

\section*{Data availability}
Data are available from the corresponding author upon reasonable request.

\begin{acknowledgements}
We thank N.~Christensen, G.~Kotliar, J.~Q.~Lin, V.~Thampy, A.~Tsvelik and W.~G.~Yin for insightful discussions, and J.~Jiang and S.~S.~ Zhang for technical support. This material is based upon work supported by the U.S. Department of Energy (DOE), Office of Basic Energy Sciences. Work at Brookhaven National Laboratory was supported by the U.S. Department of Energy, Office of Basic Energy Sciences, under Contract No.~DESC0012704. X-ray and photoemission measurements used resources at the 4-ID and 21-ID-1 beamlines of the National Synchrotron Light Source II, a U.S.\ Department of Energy Office of Science User Facility operated for the DOE Office of Science by Brookhaven National Laboratory under Contract no.~DE-SC0012704. Additional x-ray measurements used resources at 4-ID-D in the Advanced Photon Source, a U.S.\ Department of Energy (DOE) Office of Science User Facility operated for the DOE Office of Science by Argonne National Laboratory under Contract No.~DE-AC02-06CH11357. \end{acknowledgements}

\section{Competing interests}
The authors declare no competing interests.

\section{Author contributions}
H.M., T.Y., K.K., E.V., and P.D.J. performed the ARPES measurements. H.M., G.F., R.J.K., D.G.M., C.S.N., R. A.-E., T.A., I.K.R., E.S.B., and M.P.M.D performed the x-ray measurements. Y.L., G.D.G., M.O., K.K., and N.M. grew the LSCO samples and characterized their transport properties. H.M., P.D.J., and M.P.M.D. analyzed the data. H.M., J.M.T., and M.P.M.D. wrote the paper. 

\bibliographystyle{naturemag}
\bibliography{ref}

\end{document}


\title{Supplementary information: charge density waves in cuprate superconductors beyond the critical doping}

\date{\today}

\maketitle

\renewcommand{\figurename}{Supplementary figure}

\section*{Supplementary note 1: Tight binding model\label{sec_tight_binding}}
The band-dispersion shown in the main text has the form \cite{Horio2018, chang2013}
\begin{align}
\epsilon_k  = & \mu - 2 t [\cos(k_x a) + \cos(k_y a)] -4 t_1 \cos(k_x a)\cos(k_y a) \nonumber \\ 
 & -2 t_2 [ \cos(2 k_x a) + \cos(2 k_y a)] -4 t_3 [\cos(2 k_x a)\cos(k_y a) \nonumber \\ 
 & + \cos(k_x a) \cos(2 k_y a)]-4 t_4 \cos(2 k_x a)\cos(2 k_y a)  
\end{align}
where $t$ and $t_i$ ($i=1,2,3,4$) are hopping parameters. Their relative ratios are $t_1/t=-0.136$, $t_2/t =0.068$, $t_3/t=0$, $t_4/t=-0.02$, and $t =1720$~meV. Fermi-surface changes with doping are achieved by tuning the chemical potential, $\mu$, while keeping the hopping parameters unchanged.

\begin{figure}
\includegraphics[width=0.6\columnwidth]{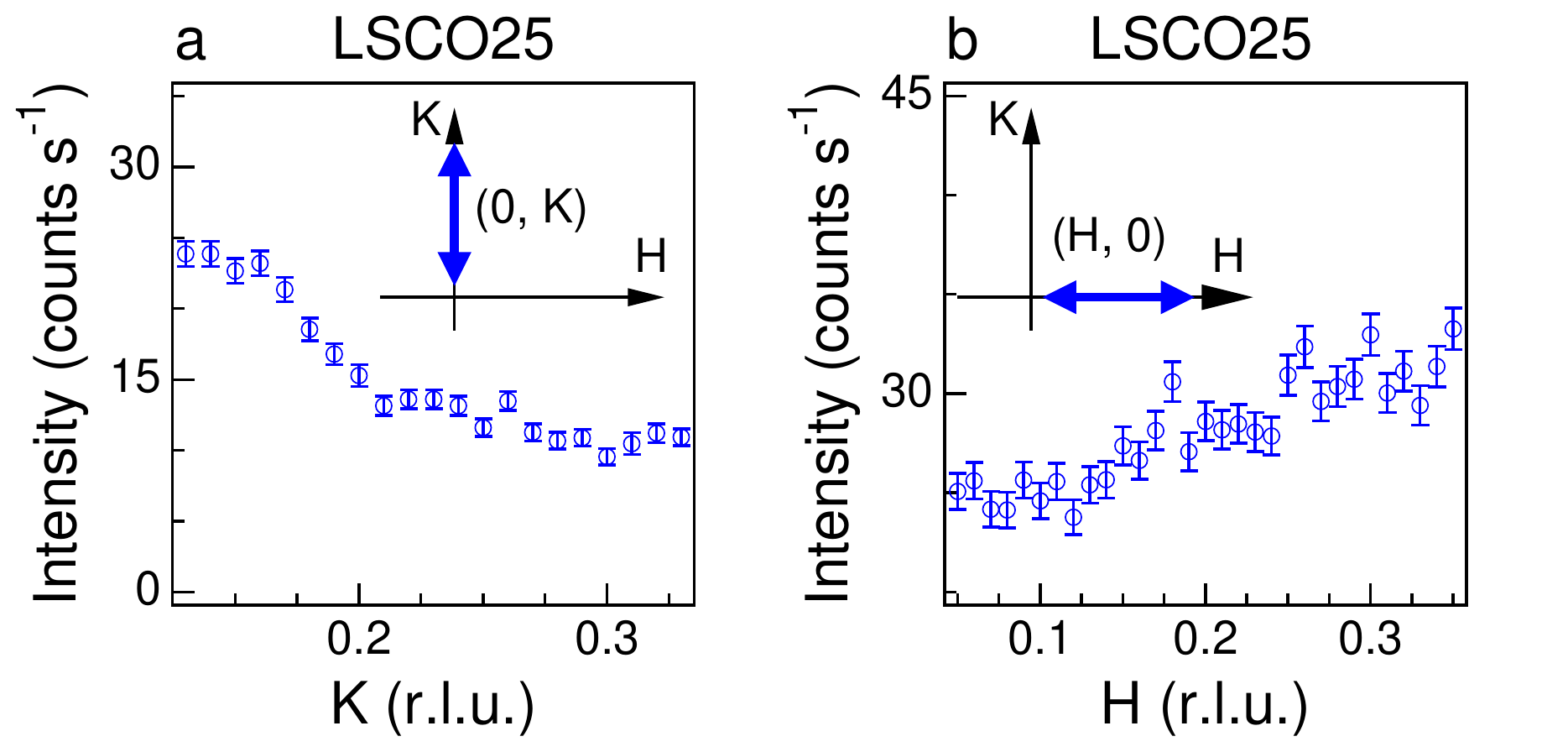}
\caption{Absence of a \gls*{CDW} peak in LSCO25. (a) and (b) shows scans along the $K$ and $H$ directions, respectively. The scans are taken at $L=12.5$. The in-plane scan trajectories are shown as insets to each panel. The data were taken at 15~K, just above $T_\text{SC}$. Errorbars are one standard deviation based on Poissonian statistics.}
\label{Fig_LSCO_diff}
\end{figure}

\begin{figure*}
\includegraphics[width=0.7\textwidth]{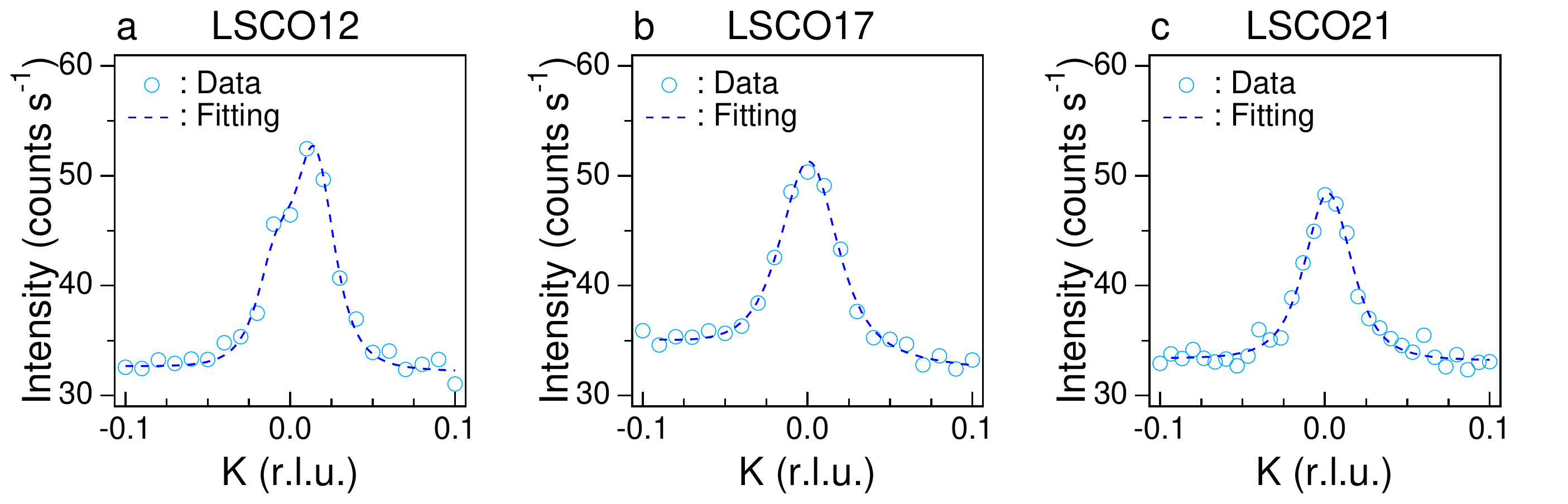}
\caption{Fitting of the \gls*{CDW} peak. Representative Lorentzian-squared fits (Eq.~\ref{EQ_Lorenzian} of the \gls*{CDW} peak perpendicular to $\bm{Q}_\text{CDW}$ near $T_\text{SC}$ are shown in (a)-(c). Note that the data for LSCO12 is composed of two peaks, as previously demonstrated \cite{Thampy2014, Wen2019}. Errorbars are one standard deviation based on Poissonian statistics.}
\label{Fig_LSCO_fit}
\end{figure*}

\section*{Supplementary note 2: Fitting of the CDW peak\label{sec_fitting}}
Following previous studies \cite{Wilkins2011, Thampy2014, Chen2016, Miao2017}, the \gls*{CDW} peaks were fitted by a Lorentzian-squared function 
\begin{equation}
I_\text{CDW}(T)= I_\text{BG}(T)+ \frac{I_0(T)}{\left(1+\left[\frac{Q-Q_0(T)}{\Gamma}\right]^2\right)^2}
\label{EQ_Lorenzian}
\end{equation}
where $I_\text{BG}(T)$ is a polynomial background, which we found to be of third order in the present dataset. The relation between $\Gamma$, (HWHM), and in-plane correlation length, $\xi_\parallel$, is given by $\text{HWHM}=\Gamma \sqrt{(\sqrt{2}-1)}$ where  $\Gamma=1/\xi_{\parallel}$. In Fig.~2 of the main text, we show representative fittings for scans along $\bm{Q}_\text{CDW}$ (longitudinal scans).  Representative fittings for scan perpendicular to $\bm{Q}_\text{CDW}$ (transverse scan) are shown in Supplementary figure~\ref{Fig_LSCO_fit}. We used two Lorentzian-squared peaks to fit transverse scans in LSCO12 due to the known splitting of \gls*{CDW} peaks are known to split at this concentration \cite{Thampy2014, Wen2019}. The HWHMs that are extracted from different transverse and longitudinal scans agree within 5\% with respect to averaged value.

\section*{Supplementary note 3: Absence of CDW signal in LSCO25\label{sec_no_CDW_LSCO25}}
Supplementary figure~\ref{Fig_LSCO_diff} shows $K$ and $H$ of LSCO25 scans near possible \gls*{CDW} wavevectors. Within our experimental uncertainty, we do not observe any \gls*{CDW} superlattice peaks at $T=15$~K $\approx T_\text{SC}$. This was observed consistently with different samples and different beamlines. The absence of a \gls*{CDW} points towards a possible connection between \gls*{CDW} correlations and non-Fermi-liquid transport as both seem to disappear concurrently (see also Appendix~V) .

\section*{Supplementary note 4: CDW intensity\label{sec_CDW_intensity}}
Assuming a weak interplanar \gls*{CDW} correlation, the integrated x-ray scattering intensity, $I^\text{int}=I_0 \  \text{HWHM}^2=I^\text{peak} (\sqrt{2}-1) \Gamma^2$, is used to estimate the magnitude of the \gls*{CDW} order parameter \cite{Thampy2013}. Former x-ray studies found that the \gls*{CDW} magnitude in LSCO12 is about four times smaller than in  La$_{1.875}$Ba$_{0.125}$CuO$_4$ and YBa$_2$Cu$_3$O$_{6.612}$ (YBCO) \cite{Thampy2013}. This is based on a direct comparison of scattering intensity in reflection using 9~keV x-rays. We note that LSCO has a shorter x-ray penetration depth of between 6.70-6.9~$\mathrm{\mu}$m (dependent on doping) compared to 8.2~$\mathrm{\mu}$m in YBCO. We remind the reader that LSCO also has fewer CuO$_2$ planes per unit volume compared to YBCO, such that approximately three times fewer CuO$_2$ planes are illumined in LSCO compared to YBCO. In this study, we use the same LSCO12 sample as previously \cite{Thampy2014}. The comparable \gls*{CDW} peak intensity and correlation length confirms that the \gls*{CDW} order parameter is of substantial size throughout the phase diagram (see Fig.~4  of the main text) and is expected to have an appreciable effect on the transport properties.

As we discussed in the main text, $I^\text{int}$ also varies with doping. Based on the fits shown in Supplementary figure~3 of the main text, we find intensity ratios of 8:7:1 at $T_\text{SC}$ for $x = 0.12$, $0.17$, and $0.21$. The magnitudes of the CDW order parameter are estimated based on data at $T=T_\text{SC}$. The appreciable reduction of $I^\text{int}$ in LSCO21 occurs close to the crossover from strange metal to Fermi liquid phase at low temperature.

\section*{Supplementary note 5: Recovery of Quasi-particle Coherence in LSCO25\label{sec_FL_LSCO25}}
The ARPES intensity map of LSCO25 at $E_F$ is shown in Supplementary figure~\ref{Fig_LSCO_ARPES}(a). In agreement with previous studies \cite{chang2013, Yoshida2001}, well-defined quasiparticles are recovered in this heavily overdoped sample as shown in Supplementary figure~\ref{Fig_LSCO_ARPES}(b)\&(c). Near $E_F$, the quasiparticle scattering rate is given by $\text{Im} \Sigma(k, \omega)=\frac{1}{2} \hbar v_F \Delta k(\omega)$, where $\text{Im} \Sigma(k, \omega)$ is the imaginary part of the self-energy, $v_F$ and $\omega$ are the Fermi velocity and the binding energy, respectively \cite{Valla1999}. $\Delta k(\omega)$ can be extracted by fitting the \gls*{MDC} with a Lorentzian function
\begin{equation}
\Gamma^{\text{MDC}}(k, \omega) = I_\text{BG}  + \frac{I_0(\omega)}{(k-k_0)^2 + (\Delta k)^2},
\label{EQ_EDC}
\end{equation}
where $I_\text{BG}(\omega)$ and $I_0(\omega)$ are constants at fixed $\omega$. Supplementary figure \ref{Fig_LSCO_ARPES}(c) shows the extracted $\Delta k(\omega)$, which can be further fitted with $a+b\omega^2$, as expected for a Fermi-liquid. One can also directly observe the emergence of a coherent quasiparticle peak the LSCO25 \gls*{EDC} that was not present when compared to an LSCO12 \gls*{EDC} at a nearby $E_F$.

\begin{figure*}
\includegraphics[width=\textwidth]{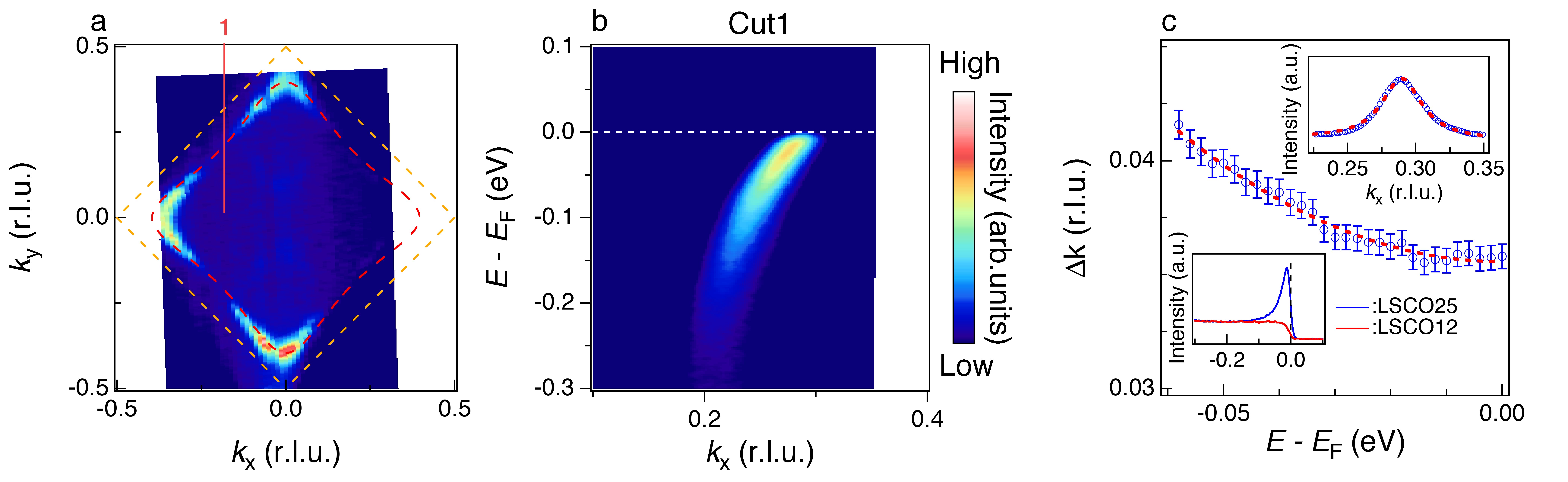}
\caption{The electronic structure of LSCO25. (a) \gls*{ARPES} intensity map at $E_F$ obtained by integrating the spectra in a $\pm10$~meV energy window with respect to $E_F$. (b) Band dispersion along the red line shown in (a). (c) Display fits to the energy and momentum distribution curves (MDCs/EDCs). The upper panel shows the MDC (blue points) and a fit to a Lorentzian-function (Eq.~\ref{EQ_EDC}) (red dashed curve) at $E_F$. The extracted MDC-width, $\Delta k$, is plotted as function of energy in the main panel of (c), which shows a $\omega^2$-dependence, which is consistent with Fermi liquid behavior. The red-dashed curve shown in the main panel of (c) is a fit of the extracted $\Delta k$ using a quadratic function, $a+b^2$. The bottom-left panel compares EDCs of LSCO12 (red) and LSCO25 (blue) at representative $k_F$, which are close in momentum space. Errorbars are one standard deviation derived from least-squares fitting.}
\label{Fig_LSCO_ARPES}
\end{figure*}

\bibliographystyle{naturemag}
\bibliography{ref}